\documentclass{kluwer}    % Specifies the document style.

\newdisplay{guess}{Conjecture}
 \usepackage{graphicx}

\begin{document}                                                                                   
\begin{article}
\begin{opening}         
\title{Rapid X-ray Variability of Seyfert 1 Galaxies}
\author{S. \surname{Vaughan}\email{sav2@star.le.ac.uk}}
\institute{X-Ray and Observational Astronomy Group, University of
  Leicester, Leicester, LE1 7RH, U.K.}
\author{A. C. \surname{Fabian}}  
\author{K. \surname{Iwasawa}}  
\institute{Institute of Astronomy, University of Cambridge, Madingley
  Road, Cambridge CB3 0HA, U. K.}
\runningauthor{Vaughan, Fabian \& Iwasawa}
\runningtitle{Rapid variability of Seyfert 1s}
%\date{August 18, 2004}

% ---------- Abstract ----------

\begin{abstract}
The rapid and seemingly random fluctuations in X-ray luminosity of
Seyfert galaxies provided early support for the standard model in
which Seyferts are powered by a supermassive black hole fed from an
accretion disc.  However, since {\it EXOSAT} there has been little
opportunity to advance our understanding of the most rapid X-ray
variability. Observations with {\it XMM-Newton} have changed this.

We discuss some
recent results obtained from {\it XMM-Newton} observations of Seyfert 1 galaxies.  
Particular attention will be given to the remarkable
similarity found between the timing properties of Seyferts and black hole
X-ray binaries, including the power spectrum and the cross spectrum
(time delays and coherence), and their implications for the physical
processes at work in Seyferts. 
\end{abstract}
%\keywords{X-rays, variability, Active galaxies}

\end{opening}           

% ---------- Main text ----------

\section{In the beginning...}  

X-ray variability appears to be ubiquitous in Active Galactic Nuclei
(AGN). 
The rapid and seemingly random fluctuations
in the X-ray luminosity of Seyfert galaxies provided early support for
the standard black hole/accretion disc model \cite{rees} by
implying compact emission regions and high luminosity densities
(\opencite{bm86}).

% --------------------------------------------------------------

\section{The {\it EXOSAT} era}  

{\it EXOSAT} (1983--1986) was the first mission to provide long
($\sim 3$ day), uninterrupted X-ray observations of Seyfert galaxies.
From these observations the X-ray 
power spectra (see \opencite{van89}) of Seyfert
galaxies were measured for the first time (\opencite{law87};
\opencite{green93}; \opencite{lp93}).   
The {\it EXOSAT} observations
showed that the power spectra of Seyferts above $\sim 10^{-5}$~Hz could be
approximated by a power-law: $\mathcal{P}(f) \propto f^{-\alpha}$
where $\mathcal{P}(f)$ is the power at frequency $f$ and $\alpha$ is
the power spectrum slope. The measured slopes from the {\it EXOSAT}
observations were typically $\alpha
\approx 1.5$. Processes such as these, which have broad-band power
spectra with more power at lower frequencies, are called ``red noise''  (see
\opencite{press}). 
It was noted early on \cite{law87} that this red
noise variability of Seyferts is similar to that observed in Galactic
Black Hole Candidates (GBHCs; \opencite{bh90}; \opencite{n99}; \opencite{mr04}),
perhaps suggesting that the same physical processes operate in these
sources that differ in black hole mass by factors of $\gsim 10^5$.

% --------------------------------------------------------------

\section{Low frequency power spectra from {\it RXTE}}

The steep slopes found in the {\it EXOSAT} power spectra required
there to be a flattening at even
lower frequencies (so that the integrated power remains finite). 
In recent years long {\it RXTE} monitoring observations 
have detected these breaks (e.g. \opencite{u02};
\opencite{mark03}; see also the article by Ian M$^{\rm c}$Hardy in
these proceedings). Below the break the slope is typically
$\alpha_{\rm lo} \approx 1$ and at frequencies above the
break the slope is $\alpha_{\rm hi} \approx 2$. 
(The {\it EXOSAT} power spectra spanned intermediate frequencies
and often measured an intermediate slope over the break.)
The breaks represent ``characteristic
timescales'' in the aperiodic variability of Seyferts and, 
significantly, appear to scale linearly with the  mass of the
central black hole: $f_{\rm break} \propto 1/M_{\rm BH}$. 

% --------------------------------------------------------------

\section{{\it XMM-Newton} results: high frequency power spectra}

\begin{figure} 
\centering
\includegraphics[width=4.0 cm, angle=270]{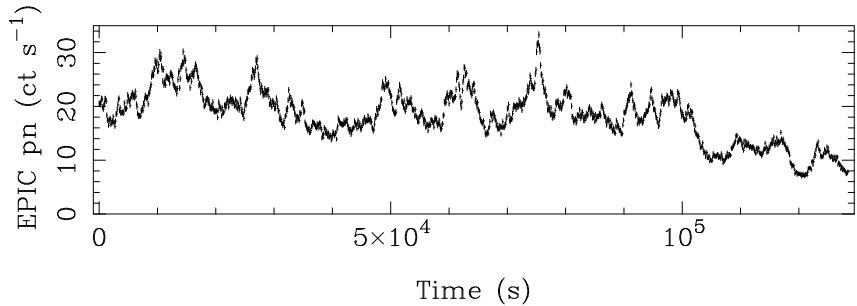}
\caption[]{
{\it XMM-Newton} light curve of Mrk 766 binned to $100$ second
resolution. 
}
\label{fig:mrk766}
\end{figure}

Until the launch of {\it XMM-Newton} \cite{jan01}
high frequency timing studies of Seyferts were not
able to substantially improve on the {\it EXOSAT} results.
{\it XMM-Newton}'s success is due to a combination of
high throughput, broad energy bandpass and long ($\sim 2$ day) 
orbit. Figure~\ref{fig:mrk766} shows an example of a broad-band
($0.2-10$~keV) light curve from a single orbit observation.

\begin{figure} 
\centering
\includegraphics[width=7.0 cm, angle=270]{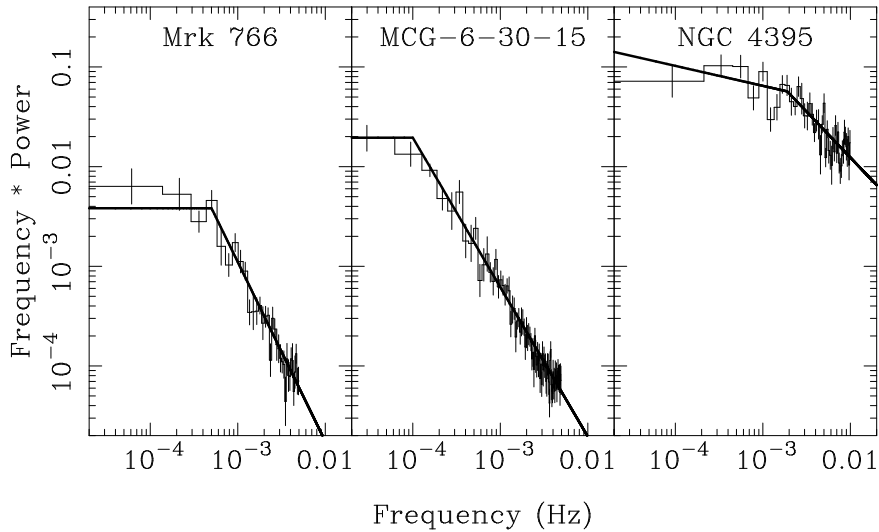}
\caption[]{
{\it XMM-Newton} power spectra of three Seyfert 1 galaxies.
Shown are the binned data (histograms) unfolded using the
best-fitting broken power law continuum model (solid line).
Note that the ordinate is in $f \times P(f)$ units
(a slope of $\alpha = 1$ would appear flat). Note
that the figure shows {\it unfolded} data, meaning the data/model
residuals multiplied by the best-fitting model. This is necessarily a
model-dependent procedure and is not advisable if sharp features are
present in the spectrum. However, for these rather broad, smooth
spectra unfolding does provide a clear, if slightly crude, impression
of the shape of the underlying spectrum free from  sampling effects.}
\label{fig:psd}
\end{figure}

Several Seyfert 1 galaxies have been studied with long  {\it
XMM-Newton} observations and yielded interesting power spectra. These
include: NGC 4051 \cite{mch04}, Mrk 766 \cite{vf03}, MCG$-$6-30-15
\cite{v03}, NGC 4395 \cite{v04} (and also Ark 564; \opencite{vig04}).
Figure~\ref{fig:psd} shows the power spectra for three of these. 
The {\it XMM-Newton} results clearly reveal similar high frequency
breaks in the power spectra (also  measured by {\it RXTE} in some
cases) but clearly show a substantial object-to-object differences in the
normalisation of the power spectrum (which describes the overall
variability amplitude).

These {\it XMM-Newton} observations have also demonstrated
the energy dependence of the power spectrum. Above the
break frequency the slope $\alpha_{\rm hi}$ tends to be
steeper at lower energies. This is clearly observed in
MCG$-$6-30-15 \cite{v03} and NGC 4051 \cite{mch04} but
is not constrained by the other observations. 
This energy dependence was also measured in NGC~7469
using an intensive {\it RXTE} monitoring campaign
\cite{np01}.

% --------------------------------------------------------------

\section{{\it XMM-Newton} results: high frequency cross spectrum}

In addition to the energy dependence of the power spectrum,
the excellent quality {\it XMM-Newton}~ light curves have
allowed the cross spectrum to be investigated in several
Seyfert 1s for the first time. Prior to {\it XMM-Newton}
only \inlinecite{p01} had measured the cross spectrum
for a Seyfert (NGC~7469 using {\it RXTE}).

The cross spectrum compares the variations in one
band with those in another as a function of frequency.
The amplitude of the cross spectrum gives the
coherence \cite{vn97} while the argument gives the
phase lag (time delay; \opencite{n99}). 
The coherence quantifies any (linear) correlation between
the variations in the two bands, irrespective of any 
time delays. 
The observations typically show high coherence
at the lowest frequencies (i.e. strong correlation) with 
a decrease at higher frequencies which implies there are
independent variations between the two bands occurring on 
short timescales
(\opencite{v03}; \opencite{mch04}; \opencite{v04}).
At low frequencies, where the coherence is high, the data also
exhibit small time delays, with the soft leading the hard
variations (\opencite{v03}; \opencite{mch04}). The magnitude
of the time delay decreases with increasing frequency (although
the functional form of the relation is poorly constrained).

% --------------------------------------------------------------

\section{Summary of results}

{\it XMM-Newton} has already made significant progress
towards improving our understanding of the high
frequency variability of Seyfert galaxies. The 
timing studies have revealed:

\begin{itemize}

\item
Similar broken power
spectra in Seyferts but object-to-object differences
in normalisation (variability amplitude) and high frequency slope.

\item
Energy-dependent high frequency power spectrum slope
(steeper at lower energies).

\item
High coherence at low frequencies, falling off
at high frequencies.

\item
Small ($\Delta T \sim 0.01/f$) soft-to-hard time delays

\end{itemize}

% --------------------------------------------------------------

\section{Comparison with GBHCs}

\begin{figure} 
\centering
\includegraphics[width=8.0 cm, angle=270]{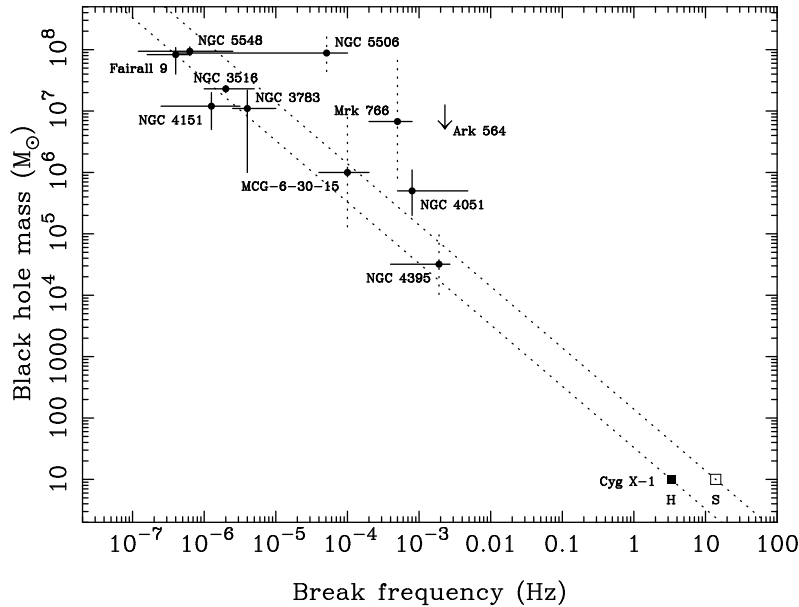}
\caption[]{
The $M_{\rm BH} - f_{\rm br}$ relation for $11$ Seyfert galaxies.
The masses are from reverberation mapping experiments except
for the four objects marked using dotted error bars.
Also shown are typical
break frequencies for Cyg X-1 in both its low/hard (H) and high/soft (S)
states. The dotted lines show example $M_{\rm BH} \propto 1/f_{\rm br}$
relations consistent with the Cyg X-1 points.
}
\label{fig:m-f}
\end{figure}

The frequencies of the breaks in the power spectra are broadly
consistent with the long-held notion that the 
characteristic frequencies should scale as $\propto 1/M_{\rm
  BH}$ right down to stellar mass black holes. 
Figure~\ref{fig:m-f} shows the available data for 11 Seyferts.
Although the uncertainties are rather large, the data
seem consistent with an extrapolation of the $ f_{\rm br} \propto
1/M_{\rm BH}$ relation from the well-studied GBHC Cygnus X-1
(\opencite{bh90}; \opencite{cui}; \opencite{n99}; \opencite{mr04}).
Note that the break frequency measurements come from a combination of 
{\it XMM-Newton} and {\it RXTE} observations. Long {\it XMM-Newton}
observations are sensitive to breaks in the range $\sim 10^{-4} -
10^{-2}$~Hz while the {\it  RXTE} monitoring campaigns are sensitive to
breaks at lower frequencies.

The connection between Seyferts and GBHCs is reinforced by the
similarity between their cross spectra. It is well known
that GBHCs show highly coherent variations at low frequencies with
the coherence fading away at the highest frequencies plus
frequency dependent time lags similar to those measured in Seyferts
(\opencite{n99}; \opencite{mr04}). These all argue for 
a common mechanism responsible for producing the X-ray variability in
Seyferts and GBHCs. 

One may ask what advantage is gained by studying Seyferts
in X-rays if GBHCs operate with the same physics but provide
much higher quality data? One answer is that Seyferts can
in fact provide data that are in some senses better than that from
GBHCs for studying the highest frequencies. For example, comparing
MCG$-$6-30-15 and Cygnus X-1 we see the timescales are longer by $\sim
10^5$ in the Seyfert, but the X-ray flux is smaller by $\sim 10^3$. 
This means that {\it per characteristic timescale} the Seyfert
provides $\sim 10^2$ more photons! Of course, GBHC enthusiasts can
argue that GBHCs reclaim much of
their advantage even here because one can always
observe many ($\sim 10^5$) more samples of a given timescale  in a fixed
amount of observing time, even if many less photons are recorded per timescale.
Even so the power spectrum of MCG$-$6-30-15
could be measured up to $\sim 5 \times 10^{-3}$~Hz using the {\it
  XMM-Newton} data. This is equivalent to probing $\sim 500$~Hz in
Cygnus X-1, a challenge for even the best {\it RXTE} observations
\cite{rev}. 

% --------------------------------------------------------------

\section{Implications for the emission processes}

The X-ray emission mechanism operating in Seyfert galaxies (and GBHCs)
is usually
thought to be inverse-Compton scattering. In the simplest
models harder photons are expected to lag behind the
softer photons due to the larger number of scatterings required to
produce harder photons; the delay should be of order the
light-crossing time of the corona. 
The direction and magnitude of the observed time lags in Seyfert 1s
are consistent 
with an origin in a Comptonising corona. However, if the lags are
frequency dependent (as expected by analogy with Cygnus X-1) the lags
at lower temporal frequencies would become much longer than expected
for a compact corona (see discussion in \opencite{n99}).   
In addition, some models
of Compton scattering coronae predict the high frequency PSD should be
steeper for higher energy photons, due to the high-frequency
fluctuations being washed out by multiple scatterings \cite{nv96},
contrary to the observations. 
Alternatively, the time delay between soft and hard bands could be due
to the spectral evolution of individual X-ray events \cite{pf} or 
propagation of accretion rate variations through a extended 
emission region \cite{kotov}.

% ---------- Acknowledgements ----------

\acknowledgements
Based on observations obtained with {\it XMM-Newton}, an ESA science
mission with instruments and contributions directly funded by ESA
Member States and the USA (NASA).
SV and KI are supported by the UK PPARC. ACF is supported
by the Royal Society. 

% ---------- References ----------

\end{article}
\end{document}